
\input phyzzx
\overfullrule=0pt
\voffset=0.0in
\hoffset=0.0in
\line{\hfill BROWN-HET-912 }
\line{\hfill SUTPD/12/93/72  }
\line{\hfill June 1993/KHOR 72}
\vskip1.5truein
\titlestyle{CUSP ANNIHILATION ON ORDINARY COSMIC STRINGS}
\bigskip
\author{M. Mohazzab\foot {On leave
from Department of
Physics, Sharif University of Technology, Tehran, Iran}}
\centerline{{\it Department of Physics}}
\centerline{{\it Brown University, Providence, RI 02912, USA}}
\bigskip
\abstract

 The order of magnitude of energy emission from cusps to light bosons on
ordinary cosmic strings is calculated perturbatively. The analysis
 is applicable to both closed string loops and long cosmic strings. The
perturbative result obtained here is much less than what is found by
non-perturbative approximations.

\endpage

{\bf\chapter{Introduction}}

Cosmic strings are linear topological defects formed during
a phase transition in the early universe [1]. They may have
 an important role in structure formation if the mass per
length parameter,$\mu $ satisfies $G\mu \simeq 10^{-6}$ [2]. This value is
consistent with the scale of the symmetry breaking in grand
 unified theories (GUT).

 Cosmic strings emit  gravitational [3] and non-gravitational [4,5,6]
 radiation. It has been shown, however, that gravitational radiation
 dominates [4,5,6].

The non-gravitatioal radiation of cosmic strings can be induced by
 oscillations of  ordinary cosmic string [4,5,6] or superconducting cosmic
string [7] and by the phenomenon of cusp annihilation [6,8].

  An important contribution to the non-gravitational radiation is  cusp
annihilation. Cusps are region of cosmic strings where the string doubles back
onto  itself, the tangent angle to string changes by $180$ degrees and its
velocity is near that of light. Cusps can be formed on long strings [9],
 as well as on loops[13].

 Here we calculate the order of magnitude of particle production at cusps  on
ordinary cosmic strings by perturbative field theoretical methods and compare
the result with the naive non perturbative assumptions[6].
 In this work we generalize the field theoretical calculations of [4,6] to more
general string configurations (closed and long string), focusing on the cusp
region. We will see that the amount of energy predicted by this work is much
smaller than what was obtained by  previous non-perturbative approximations for
cusp annihilation.
As a result, the background gamma-ray attributed to cosmic strings would
 be much smaller than what was predicted by  using the other approximations
[9,14].

  For a global $U(1)$ string the action, in terms of its constituting field
$\phi $, is

$$S= \int d^4x\sqrt{-g}[{1\over 2}\partial_\mu \phi
\partial^{\mu}\phi - f(\vert \phi \vert^2- \sigma^2)^2],\eqno\eq$$
\noindent
where $\phi $ is a complex scalar field and f is the self coupling constant.
Assuming that the field configuration corresponds to a string with world sheet
 $\chi^{\mu} (s, \tau)$ in four dimensional space-time, where $s ,\tau$ and
$h_{ij}$ are its world sheet
coordinates and metric, the above field theory action reduces [11] to the Nambu
action for the world sheet:
$$S = \mu \int d^2 s \sqrt{h} \partial_i \chi_{\mu} (s, \tau
) \partial_j \chi^{\mu} (s, \tau ) h^{ij} \eqno\eq$$
where  $i$ and $j$ are world sheet indices. This action to a good accuracy
describes the motion of the string world sheet provided

$${w\over R} \ll 1\eqno\eq$$
where $w$ and $R$ are the width and
curvature radius of the string, respectively. The width $w$ of the string is
given in terms of $\mu$ by $w \sim \mu^{-1/2}$. In case the equation (1.3) is
not valid the action (1.2) should be corrected by higher order terms [12].

  By the equations of motion derived from the action $(1.2)$ it can be shown
that cusps, or the points where $\dot \chi =1$ and $\chi' =0$ can be formed on
loops [13] or long strings[9].

{\bf\chapter {The field theoretical calculation of cusp annihilation}}

  Cusps are regions where the cosmic strings double back onto themselfes.
 Hence, there is no topological constraint which prevents the strongly
correlated string field configuration from decaying into  outgoing jets of
particles which can be fermionic or bosonic.  As a toy model, inspired by the
field potential of the cosmic string we consider  the interaction lagrangian
$${\cal L}_I = f \tilde \phi^2 \Psi^2\eqno\eq$$
where $\Psi$ is the bosonic field of the outgoing pair of
particles,
$\tilde \phi $ is the real part of the higgs field of string expanded
 around the real vacuum $\vert \langle \phi \rangle \vert = \sigma $,
and $f$ is the cupling constant.

For a long string, we consider a line of string in spatial
$z$ direction with small wiggles on it and asign a quantum
state $\vert S\rangle $ to its
configuration.  The state could be interpreted as a coherent state of
higgs field configuration.

The initial state can be $\vert S\rangle $ and the final state $
\vert S\rangle {\vert \psi (k_1 )\psi (k_2)\rangle }$, where $\vert
\psi (k) \rangle $
represents the bosonic outgoing state with momentum $k$.

The $S$-matrix element
$$S_{fi} = {\langle S,\psi (k_2) \psi (k_1)}\vert {S\rangle } \eqno\eq$$
 can be written using the $LSZ$
construction as
$$S_{fi} = f \int d^4 x e^{i(k_1 + k_2 )\cdot x} \, {\langle S} \vert :
\tilde \phi^2 : \vert {S\rangle  }\eqno\eq$$
where :: denotes normal ordering.

Calculating (3) requires some knowledge of $\tilde \phi$.  There is,
 however, no
solution in closed form for the cosmic string field configuration.
   Here, we use the following approximation [4,6]
$${\langle S} \vert : \tilde \phi^2 (\chi ) :\vert S\rangle \simeq \sigma^2
w^2\,\int
ds \vert \chi' \vert^2
\delta^3  (\bf x - \bf \chi (t, s))\eqno\eq$$
where $\sigma $ and $w$ are the value of $\phi $ at the center of the string
and its thickness respectively.
${\bf \chi }(t, s)$ represents the world sheet of the traveling wave [10] along
the string.

Now the integral (2.3) can be written as
$$\eqalign{S_{fi}  & = f \int d^4 x \, \int d s
 \vert \chi' \vert^2 e^{ikx} \sigma^2 w^2\delta ( {\bf x} - {\bf \chi }(t,
s))\cr
& = f \sigma^2 w^2\int dt \, ds \, \vert \chi' \vert^2
e^{i(Et-{\bf k}\cdot {\bf \chi }(t, s))} } \, \eqno\eq$$

The main contribution to (2.5) is from the cusp vicinity where the phase
$\Phi $ is stationary, namely
$$\eqalign{ \dot \Phi (t,s) & = 0\cr
\Phi'  (t,s) & = 0 }\eqno\eq$$
where $\cdot$ and $'$ mean the derivatives with respect to time and
$s$, respectively.

  We use the SPG[6] result to expand the determinant and phase of
the integrand (2.5) around the cusp site:

 $$\chi =\left(\matrix{t\cr t-{1\over 2}(\alpha^2 +\beta^2 +\gamma^2
)({1\over 3}(t^3 +s^2t))-\alpha \beta (t^2s +{1\over 3}s^3) \cr
\alpha ({1\over 2}(t^2+s^2))+\beta st +\delta ({1\over
3}t^3+s^2t)+\zeta (t^2+{1\over 3}s^3) \cr
\gamma st +\epsilon ({1\over 3}(t^3+s^2t)+\xi (t^2s+{1\over
3}s^3)\cr}\right)\eqno\eq$$
where $\alpha, \beta, \gamma ,\delta ,\epsilon ,\xi $ and $\zeta $ are
some  constants of dimension $1\over {length}$ or $1\over
{(length)^2}$ and for closed string with size $R$, approximately ${1\over R}$.
Here, we assume $\alpha \approx \beta \approx
\gamma $.
 Note that the expansion (2.7) is approximately valid while $\vert t \vert \&
\vert s \vert \ll {1\over \alpha }$.

 The parameter $\alpha, $ with dimension ${1\over {length}}$, is proportional
to the size of the incoming traveling wave [10] on strings. For closed strings
with typical length $R$ we have [6, 8] $\alpha^{-1}\sim R$ and for long string
with incoming traveling wave size $l$, we have $\alpha^{-1} \sim l$.

Plugging the expansion (2.7) into the stationary phase condition (2.6) it can
easily be seen that
$k_x \simeq E$ . Hence, by the mass shell condition $k_y  \&   k_z \approx 0$
near the cusp . More precisely the condition

  $$k_z , k_y \ll E^{2\over 3}\alpha^{1\over 3} \eqno\eq$$
can be obtained from (2.6) and the relation (2.12)
 and therefore the maximum angle between the two vectors
$\bf k_x$ and $\bf k_y$ can be approximated as
  $$\theta_{max} \sim {k_y \over k_x }\sim E^{-{1\over 3}}\alpha^{1\over
3}.\eqno\eq$$
  By substituting $z=s$, the phase in the integrand (2.5) will take the form
 $$\Phi \simeq [{1\over 2}\alpha^2 {1\over 3}(t^3+z^2t)-\alpha^2 (t^2z+{1\over
3}z^3)]E.\eqno\eq$$
  The considerable contribution to the integral (2.5) comes from the region
where
   $$\Phi \lsim 1\eqno\eq$$
or
 $$t_{max}, z_{max} \leq ({1\over {\alpha^2 E}})^{1\over 3}.\eqno\eq$$
 Thus,  $t_{max}$ and $z_{max}$ will be the upper bounds for the integration.

 Now the S-matrix (2.5) can be written as

   $$S_{fi} =f\sigma^2 w^2
\int^{t_{max}}_{-t_{max}} dt\int^{z_{max}}_{-{z_{max}}}dz \Delta^2
e^{iE\Phi }\eqno\eq$$
where $\Delta^2$ is approximately
$$\Delta^2 \simeq \alpha^2 (z^2+t^2)+\alpha^4 (z^2t^2+z^4+t^4)\eqno\eq$$
where we have considered the terms in the integral that do not vanish
and absorbed all the constants in $\alpha $.
  The total energy emitted can be found by working out the following
integral

$$\eqalign{\tilde {E} & = \int {d^3 k_1\over (2\pi )^3 k_1^0} \, {d^3
k_2\over (2\pi )^3 k_2^0} \, \vert S_{fi} \vert^2 E\cr
& = f^2\sigma^4 w^4\int {d^3 k_1\over (2\pi )^3 k_1^0} \, {d^3 k_2\over (2\pi
)^3
k_2^0} \, (k_1^0 + k_2^0 )\bigg\vert \int dt\int dz \Delta^2e^{i\Phi
}\bigg\vert^2 }\eqno\eq$$
The upper limit for $(k_1^0 + k_2^0) $ is the scale of the  energy of the
cosmic string i.e $\sigma $. More precisely this  cutoff can be figured out
from the action (1.1), by considering the mass of  particles i.e. $f^{1\over 2}
\sigma $ [6,15].
Therefore the compton wavelength will be $d_{c} \sim {1\over {f^{1\over
2}\sigma }}$, where $f$ is the coupling constant and $f\lsim 1.$
The energy cutoff will, consequently, be the amount of energy in the string
within a distance $d_c$, i.e.

$$E_{cut-off}\sim {1\over d_c}\sim f^{1\over 2}\sigma \eqno\eq$$
An upper bound to (2.15) can be founded
by approximating $d\Omega$ using (2.9)

$$d\Omega \sim \theta_{max}^2 \sim E^{-{2/3}}\alpha^{2\over 3}\eqno\eq$$
Therefore, the upper limit to the energy of ejected particles from a
cusp, $\tilde{E}_{max}$, will be

$$\eqalign{ \tilde{E}_{max} & = {1\over (2\pi )^6} \, {\sigma^4 w^4
f^2\over {\alpha^{4\over 3}}}\int_{k_{1}^{0} = E_{min}}^{f^{1\over 2}\sigma }
k_{1}^{0} dk_1^0 \int_{k_{2}^{0} = E_{min}}^{f^{1\over 2}\sigma -k_{1}^{0}}
k_2^0 dk_2^0 (k_1^0
+ k_2^0 ) d\Omega_1 d\Omega_2 \, (k_1^0 +
k_2^0 )^{-8/3}\cr
& k_1^0 , k_2^0  \in [ E_{min}, f^{1\over 2}\sigma ]} \eqno\eq$$

 For $E_{min}\ll \sigma$, eq.(2.17) can be written as
$$\tilde{E}_{max} \sim  f^2\, w^4\sigma ^4
{1\over 2}(f^{1\over 2}\sigma -{3\over 2}E_{min})\eqno\eq$$
 Therefore the total energy of the emitted pairs can be easily found by
 substituting $\sigma w = 1$ and neglecting $E_{min}$, namely

$$E_{tot} ={1\over 2}f^{5\over 2}\sigma \eqno\eq$$
As a result the predicted amount of energy emitted by cusp annihilation, in
this perturbative calculation, is proportional to the energy scale of symmetry
breaking of the cosmic strings.

In conclusion, considering the absence of back reactions due to gravitational
radiation, we have calculated to an order of magnitude, the maximum amount of
energy that the light bosons can take when cusps on cosmic strings annihilate.
These bosons can further decay to bursts of gamma-rays and contribute to the
gamma-ray background [9,14].

The maximum energy is propotinal to $\sigma $, the scale of symmetry breaking
of the cosmic strings. Using equation (2.8) and the fact that $\alpha \ll
\sigma$ (i.e. the size of the incoming traveling wave is much bigger than the
thickness of the cosmic string), it is easy to show that
       $$\sigma \ll E_{cusp}\eqno\eq $$
where $E_{cusp}$ is the energy in the cusp region, namely
  $$E_{cusp}=\mu l_{cusp}=\mu \sigma^{-{1\over 3}} \alpha^{-{2\over 3}}$$
where $l_{cusp}$ is the length of the cusp(overlap) region.
Therefore the maximum amount of energy predicted by the perturbative
calculations is much less than the results obtained by non-pertubative
approximations [6].  As a result the background gamma ray predicted by this
work is much smaler than what was determined in the previous calculations[9,14]

{\bf ACKNOWLEDGMENTS}

 I wish to thank Robert H. Brandenberger for many useful discussions and
insights. The author is grateful to Brown University for hospitality and to the
Ministry of Culture and Higher Education of Iran for financial support.

\endpage

\noindent{{\bf References}}

\pointbegin
A. Vilenkin, {\it Phys. Rep.} {\bf 121}, 263, (1985).
\point
L. Perivolaropoulos, R. H. Brandenberger and A. Stebbins, {\it Phys. Rev.}
 {\bf D41}, (1990), 1764.
\point
R. R. Caldwell and B. Allen,{\it Phys. Rev.} {\bf D 45} (1992) 3447.
\point
M. Srednicki and S. Theisen, {\it Phys. Lett.} {\bf B 189}, (1987), 397.
                                                \point
A. Vilenkin, {\it Phys. Rev.} {\bf D 30} (1984) 2046
\point
R. H. Brandenberger, {\it Nucl. Phys.}{\bf B293},(1987), 812.
\point
E. Witten, {\it Nucl.Phys.}{\bf B 242},(1985), 557.
\point
D. N. Spergel, T. Piran and J. Goodman, {\it Nucl. Phys.} {\bf B 291},
(1987) 847.
\point
M. Mohazzab and R. H. Brandenberger, BROWN-HET-892 (SUTPD/93/72/7),
{\it Int. J. Mod. Phys.} {\bf D 2},  (1993).
\point
Vachaspati and T. Vachaspati, {\it Phys. Lett.} {\bf B 238}, (1990)
41; D. Garfinkle and T. Vachaspati, {\it Phys. Rev.} {\bf D 42}, (1990)
1960.
\point
D. Foerster, {\it Nucl. Phys.} {\bf B 81} (1974) 84;
N Turok, {\it Proc. 1987 CERN/ESO Winter School on Cosmology and Particle
Physics, Erice} (Singapore: World Sceintific).
\point
R. Gregory, {\it Phys. Lett.} {\bf B206}, (1988), 199.
\point
T. W. B. Kibble and N. Turok, {\it Phys. Lett.} {\bf B116}, (1982), 141.
\point
J. H. MacGibbon and R. H. Brandenberger, {\it Nucl. Phys.} {\bf B 33},
(1990), 153;
J. H. MacGibbon and R. H. Brandenberger, {\it Phys. Rev. } {\bf D47}, 2283,
(1993);P. Bhattacharjee, {\it Phys. Rev. }{\bf D 40}, (1989) 3968.
\point
R. H. Brandenberger, private communication.
\end